\documentclass{optica-article}

\journal{opticajournal} 

\articletype{Research Article}

\usepackage{siunitx}
\usepackage{url}
\usepackage[version=4]{mhchem}
\usepackage{float}
\usepackage{graphicx}
\usepackage{subcaption}
\usepackage{fix-cm}
\usepackage{booktabs}
\usepackage{xcolor}
\usepackage{dcolumn}
\usepackage{bm}
\usepackage{multirow}
\usepackage{svg}
\usepackage{appendix}
\usepackage{mathtools}
\usepackage{makecell}
\usepackage{amsmath}
\usepackage{accents}
\usepackage{braket}
\usepackage{tikz}

\graphicspath{{./figures/}}

\begin{document}

\title{Joint spectral characterization of SPDC photon pairs near 2 \textmu m in (Al)GaAs-on-insulator waveguides}

\author{Alexandre Z. Leger\authormark{1,*}, Emil Z. Ulsig\authormark{2,3}, Samuel E. Fontaine\authormark{4}, Dileep V. Reddy\authormark{5}, Eric J. Stanton\authormark{3,6}, Lynden K. Shalm\authormark{5}, Richard P. Mirin\authormark{5}, Martin J. Stevens\authormark{5}, J. E. Sipe\authormark{4}, Nicolas Volet\authormark{2,3}, and Deny R. Hamel\authormark{1}}
\address{\authormark{1}Département de physique et d'astronomie, Université de Moncton, Moncton, New Brunswick E1A 3E9, Canada\\
\authormark{2}Department of Electrical and Computer Engineering, Aarhus University, Aarhus, Denmark\\
\authormark{3}QVersion, Aarhus, Denmark\\
\authormark{4}Department of Physics, University of Toronto, 60 St.\ George Street, Toronto, Ontario M5S 1A7, Canada\\
\authormark{5}National Institute of Standards and Technology, Boulder, CO 80305, USA\\
\authormark{6}EMode Photonix, Boulder, CO 80301, USA}

\email{\authormark{*}eal2753@umoncton.ca}


\begin{abstract*}
Integrated photon-pair sources are a core component of chip-based quantum computing, communication, and metrology. Although such sources have been demonstrated at conventional telecom wavelengths, the \qty{2}{\micro\meter} band remains comparatively less explored, despite offering advantages for free-space quantum communication, low-loss transmission in emerging fiber networks, and integration with silicon photonic platforms. In this paper, we demonstrate spontaneous parametric down-conversion (SPDC) in straight \ce{GaAs}- and \ce{AlGaAs}-on-insulator waveguides. This platform offers strong second-order nonlinearity and geometry-tunable dispersion, which are advantageous for efficient on-chip pair generation. Measurements of the joint spectral intensity and heralded second-order correlation function show broadband emission around \qty{2}{\micro\meter} with strong spectral anti-correlations. To our knowledge, this is the first direct joint-spectral characterization of an integrated SPDC source in this wavelength regime.
\end{abstract*}

\section{Introduction}

The development of on-chip photon-pair sources operating beyond conventional telecom wavelengths enables several emerging quantum technologies. The \qty{2}{\micro\meter} band, in particular, offers favorable atmospheric transmission windows with
reduced Rayleigh scattering and lower solar background than shorter near-infrared wavelengths, benefiting daylight free-space quantum key distribution~\cite{liaoSatellitetogroundQuantumKey2017,prabhakarTwophotonQuantumInterference2020}. It is also compatible with low-loss transmission in hollow-core fibers under development for quantum networks~\cite{zhangHollowCoreNANFHighFidelity2024}, and with the silicon photonic platforms maturing at these wavelengths~\cite{linMidinfraredIntegratedPhotonics2018,roelkensSiliconBasedPhotonicIntegration2014}. Photon pairs in this band also enable quantum spectroscopy and metrology, where many molecular resonances are inaccessible from the telecom band. These applications require bright, well-characterized on-chip sources that can be monolithically integrated with photonic circuits.

Previous work on photon-pair generation at \qty{2}{\micro\meter} has used spontaneous four-wave mixing in silicon waveguides~\cite{signoriniSiliconSourceHeralded2021} and SPDC in bulk crystals~\cite{prabhakarTwophotonQuantumInterference2020}, with integrated $\chi^{(2)}$ SPDC recently reported in thin-film periodically poled lithium niobate nanophotonic devices~\cite{williamsUltrashortPulseBiphoton2024}. While ultrabroadband SPDC in thin-film lithium niobate spans from the telecom band into the \qty{2}{\micro\meter} range~\cite{javidUltrabroadbandEntangledPhotons2021}, the joint spectral intensity of an integrated SPDC source in this band has not been directly measured. 

Among candidate platforms, \ce{GaAs}~\cite{babouxNonlinearIntegratedQuantum2023} combines a large second-order nonlinearity, a strong electro-optic effect suitable for state transformation~\cite{dietrichGaAsIntegratedQuantum2016}, and a mature III--V fabrication ecosystem that supports the heterogeneous integration of superconducting single-photon detectors~\cite{mcdonaldIIIVPhotonicIntegrated2019}. The addition of aluminum widens the bandgap to suppress two-photon absorption, and the geometry-tunable dispersion, set by the epitaxially grown film thickness and the lithographically defined ridge width, allows precise phase matching of efficient nonlinear processes~\cite{ulsigEfficientWidelyTunable2024,stantonEfficientSecondHarmonic2020}; these capabilities have recently been used to demonstrate bright SPDC at \qty{1550}{\nano\meter}~\cite{plackeTelecomBandSpontaneousParametric2024} and heterogeneous integration with silicon-on-insulator quantum circuits~\cite{lazzariBiphotonStateGeneration2026}. 

Extending this platform to \qty{2}{\micro\meter} introduces experimental challenges for characterizing the emitted pairs. In particular, measuring the joint spectral intensity (JSI), which provides important information about the purity and indistinguishability of the generated photons~\cite{zielnickiJointSpectralCharacterization2018}, is limited by the low efficiency and high dark-count rates of infrared detectors as well as the high propagation losses of standard fibers at these wavelengths.

In this work, we demonstrate and characterize photon pairs generated by SPDC in straight \ce{GaAs}- and \ce{AlGaAs}-on-insulator waveguides phase-matched around \qty{1960}{\nano\meter}. We extend the characterization of integrated SPDC sources into the infrared using custom superconducting nanowire single-photon detectors (SNSPDs) optimized for this wavelength, and, to our knowledge, we present the first measurement of the JSI of an integrated SPDC source in this regime, obtained directly by time-of-flight spectroscopy in a dispersive fiber~\cite{zielnickiJointSpectralCharacterization2018}. We further quantify the down-conversion efficiency, the propagation losses of the pump and signal--idler modes, and the heralded second-order correlation function $g^{(2)}_\mathrm{H}$, and we develop a loss-inclusive numerical model that reproduces the observed broadband spectra and informs future device designs.

\section{Fabrication}

\begin{figure}[!ht]
    \centering
    \includegraphics[width=0.55\linewidth]{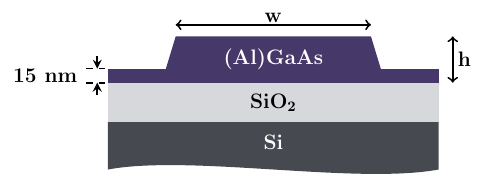}
    \caption{Cross-section of the (Al)GaAs-on-insulator ridge waveguide geometry. The (Al)GaAs core and slab rest on a \ce{SiO2} layer above a silicon substrate, with air upper cladding. The sidewall angle, estimated from a scanning electron micrograph, is $\sim$17$^{\circ}$~\cite{ulsigEfficientWidelyTunable2024}.}
    \label{fig:ModeProfiles}
\end{figure}

The devices, shown in Fig.~\ref{fig:ModeProfiles}, are rectangular ridge waveguides designed for modal phase matching. Full design and fabrication details are provided in the supplementary material of Ref.~\cite{ulsigEfficientWidelyTunable2024}; briefly, the (Al)GaAs films are grown by molecular beam epitaxy on GaAs substrates, directly bonded to a thermally oxidized silicon wafer, after which the growth substrate and an etch-stop layer are removed by selective wet etching~\cite{ulsigEfficientWidelyTunable2024}.

The (Al)GaAs-on-insulator ridge structures use width and height tuning to achieve phase matching by equalizing the effective indices at the pump, signal, and idler wavelengths. The GaAs design reproduces a previously demonstrated SHG geometry ($\approx \qty{40}{\per\watt}$)~\cite{stantonEfficientSecondHarmonic2020}, and the \ce{Al_xGa_{1-x}As} devices follow the same approach with $x{=}0.15$ to increase the bandgap and access longer wavelengths. Phase-matching dimensions were identified with a finite-difference mode solver (EMode \cite{EMode}), and devices were fabricated across a parameter window around the simulated geometry.

For characterization, sets of GaAs and AlGaAs waveguides were fabricated in parallel under identical conditions. AlGaAs chips from the same batch were also used in prior work to demonstrate frequency conversion, including difference-frequency generation to telecom wavelengths \cite{ulsigEfficientWidelyTunable2024}. The waveguides are phase-matched to convert a \qty{980}{\nano\meter} pump photon in the transverse magnetic (TM) mode into two transverse electric (TE) mode photons, centered at \qty{1960}{\nano\meter}.

\begin{figure}[t]
    \centering
    \includegraphics[width=0.95\linewidth]{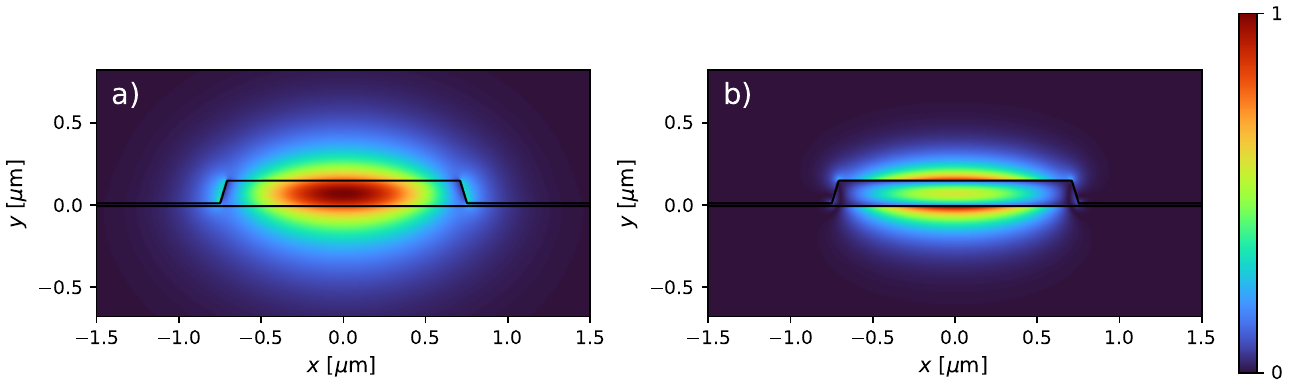}
    \caption{Magnitude of the dominant electric-field component (a) for the fundamental TE$_{00}$ signal/idler mode and (b) for the TM$_{00}$ pump mode,
    calculated for the AlGaAs waveguide geometry using
    the numerical model described in Section~\ref{section:numerical}.
    Similar mode profiles are obtained for GaAs.}
    \label{fig:ModeFields}
\end{figure}

\section{Experiments}

\subsection{SPDC efficiency, loss, and second-order correlation measurements}\label{subsec:efficiency}

Light was collected from the output facet of the chip with a lensed fiber, and residual pump light was filtered out before the signal--idler photons were routed in fiber to the detectors. The photon pairs were detected using niobium-titanium nitride (\ce{NbTiN}) SNSPDs. 

The meanders were \qty{500}{\nano\meter}-wide wires etched into a \qty{3.5}{\nano\meter}-thick \ce{NbTiN} film in a high-fill-factor candelabra-style pattern, providing broadband, polarization-insensitive detection~\cite{reddyBroadbandPolarizationInsensitivity2022}. They were embedded in an all-dielectric optical stack consisting of a 13-layer distributed Bragg reflector and a 2-layer top cladding optimized for absorption at wavelengths around \qty{2}{\micro\meter}. The detectors had a large active area, giving an internal geometric jitter of \qty{200}{\pico\second}. They were coupled to AR-coated SM1950 fibers and operated at \qty{850}{\milli\kelvin} in a sorption cryostat. 

\begin{figure}[tb]
    \centering
    \includegraphics[width=0.95\linewidth]{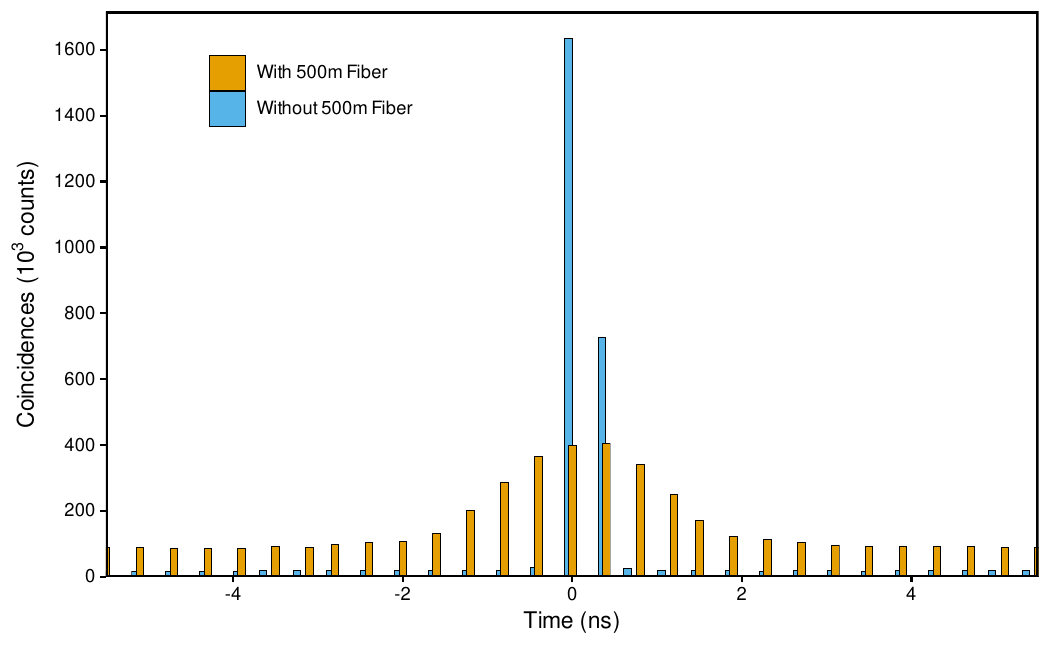}
    \caption{Coincidence histograms of photon pairs from a GaAs waveguide pumped by a continuous-wave diode laser at \qty{975}{\nano\meter}, measured with and without the \qty{500}{\meter} dispersive fiber. Peak broadening arises from the fiber chromatic dispersion used for time-of-flight spectral analysis. Without the fiber, the histogram was accumulated for \qty{90}{\second}, yielding a coincidence-to-accidental ratio of 90. The measurement with the fiber required a 20-minute accumulation, which reduced this ratio to 4; the extended accumulation time and lower ratio account for the elevated background.}\label{fig:histogram}
\end{figure}

Detector performance and signal quality were assessed by measuring the coincidence histogram of the photon pairs, shown in Fig.~\ref{fig:histogram}. The measurement without the dispersive fiber exhibits a sharp peak with a width limited by the combined timing jitter of the detectors and electronics.

We estimate the SPDC conversion efficiency by first measuring the propagation losses of the pump and signal--idler modes. Propagation and coupling losses were characterized using a cut-back method on paperclip structures on the same chip, from which we extracted per-facet pump coupling losses of $20.0 \pm 4.5$~\unit{\decibel} in GaAs chips and $23.3 \pm 3.1$~\unit{\decibel} in AlGaAs chips. Applying the Klyshko method \cite{klyshkoUseTwophotonLight1980} to the measured coincidences-to-singles ratios gives the heralding efficiencies of both detection arms. With these efficiencies and the on-chip pump power derived from the pump coupling loss estimates, we calculate the probability that an incident pump photon undergoes down-conversion into a signal--idler pair. 

The GaAs efficiency was measured with a mode-locked Ti:sapphire laser producing pulses centered at \qty{980}{\nano\meter} with a repetition rate of approximately \qty{80}{\mega\hertz}, tuned to the phase-matching wavelength; the AlGaAs efficiency was measured in a separate session with a fixed-wavelength continuous-wave (CW) pump, as availability of the tunable pulsed source was limited. All quoted pump powers are average powers. Because this fixed-wavelength source could not be tuned to the phase-matching wavelength, the AlGaAs pump was set to \qty{975.3}{\nano\meter}, which sits $\delta = 1.8 \pm 0.3$~\unit{\nano\meter} below the device's phase-matching wavelength; the uncertainty on $\delta$ is set by the precision with which the coincidence maximum locates the phase-matching wavelength, of order half the \qty{0.5}{\nano\meter} (FWHM) bandwidth of the tunable Ti:sapphire pump. This detuning reduces the AlGaAs generation efficiency relative to the phase-matched case; the model is evaluated at the same detuning for comparison (Section~\ref{section:numerical}). The pair generation probability is given by:
\begin{align}
P_\mathrm{SPDC}=\frac{C}{\eta_1\eta_2}\frac{h c}{\lambda_\mathrm{p} P_\mathrm{eff}},
\label{eq:P_SPDC}
\end{align}
where $C$ is the number of measured coincidences per second within a window of \qty{1}{\nano\second}, $\eta_\mathrm{1}$ and $\eta_\mathrm{2}$ are the efficiencies of the two heralding arms, and $\lambda_\mathrm{p}$ and $P_\mathrm{eff}$ are the pump wavelength and effective on-chip power, respectively. The latter is defined as the spatial average of the coupled pump power over the waveguide length, accounting for the exponential decay induced by propagation losses, and is given by
\begin{align}
P_\mathrm{eff} &= \frac{1}{L}\int_0^L P_\mathrm{in}\,e^{-\alpha_{\mathrm{p}} z}\,\mathrm{d}z
= \frac{1-e^{-\alpha_{\mathrm{p}} L}}{\alpha_{\mathrm{p}} L} P_\mathrm{in},
\end{align}
where $z$ is the distance along the waveguide, $P_\mathrm{in}$ is the average on-chip pump power after the input facet, estimated from the cut-back method, $L$ is the waveguide length, and $\alpha_{\mathrm{p}}$ is the linear propagation loss of the pump. For these measurements, $L=\qty{1678}{\micro\meter}$, and the corresponding $\alpha_{\mathrm{p}}$ values are reported in Table~\ref{tab:Params}. The conversion probability data are summarized in Table~\ref{tab:psdpc}.

\begin{table}[ht]
\centering
\renewcommand{\arraystretch}{1.3}
\begin{tabular}{@{} c | c | c c c | c c | c @{}}
\hline\hline
Material & \makecell[c]{$P_\mathrm{in}$ \\ (\unit{\micro\watt})} & \makecell[c]{$S_1$ \\ (\unit{\hertz})} & \makecell[c]{$S_2$ \\ (\unit{\hertz})} & \makecell[c]{$C$ \\ (\unit{\hertz})} & \makecell[c]{$\eta_1$ \\ (\unit{\percent})} & \makecell[c]{$\eta_2$ \\ (\unit{\percent})} & \makecell[c]{$P_\mathrm{SPDC}$ \\ ($\times10^{-6}$)} \\
\hline
GaAs   & $2.58^{+3.07}_{-1.42}$ & 925k  & 1015k & 41500 & 4.49 & 4.09 & $1.77^{+2.11}_{-0.98}$ \\
\hline
AlGaAs & $52.6^{+55.8}_{-27.4}$ & 1106k & 900k  & 9500  & 0.86 & 1.05 & $0.51^{+0.61}_{-0.28}$ \\
\hline\hline
\end{tabular}
\caption{Measured photon down-conversion probabilities for the best performing waveguides using the Klyshko method. $S_1$ and $S_2$ are the singles rates measured in each heralding arm. For the GaAs measurement, a pulsed Ti:sapphire laser was used as the pump source, while for the AlGaAs measurement, a CW laser was used as the pump source.} \label{tab:psdpc}
\end{table}

We also measured the heralded second-order correlation function $g^{(2)}_\mathrm{H}(\tau)$ to confirm the nonclassical nature of the emitted light and quantify the impact of multi-pair emission, where $\tau$ is the time delay between the two idler detectors.

\begin{figure}[t]
    \centering\includegraphics[width=1.0\textwidth]{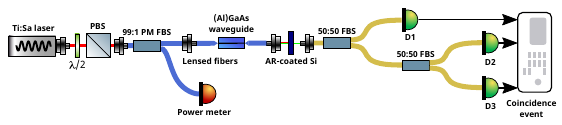}
    \caption{Setup used to measure the heralded second-order correlation function. A half-wave plate and polarizing beam splitter (PBS) control the pump power prior to coupling into the waveguide. Residual pump light is suppressed using anti-reflection (AR) coated silicon before the down-converted signal is directed to a three-detector correlation measurement.}\label{fig:g2_setup}
\end{figure}

\begin{figure}[tb]
    \centering
    \includegraphics[width=0.95\linewidth]{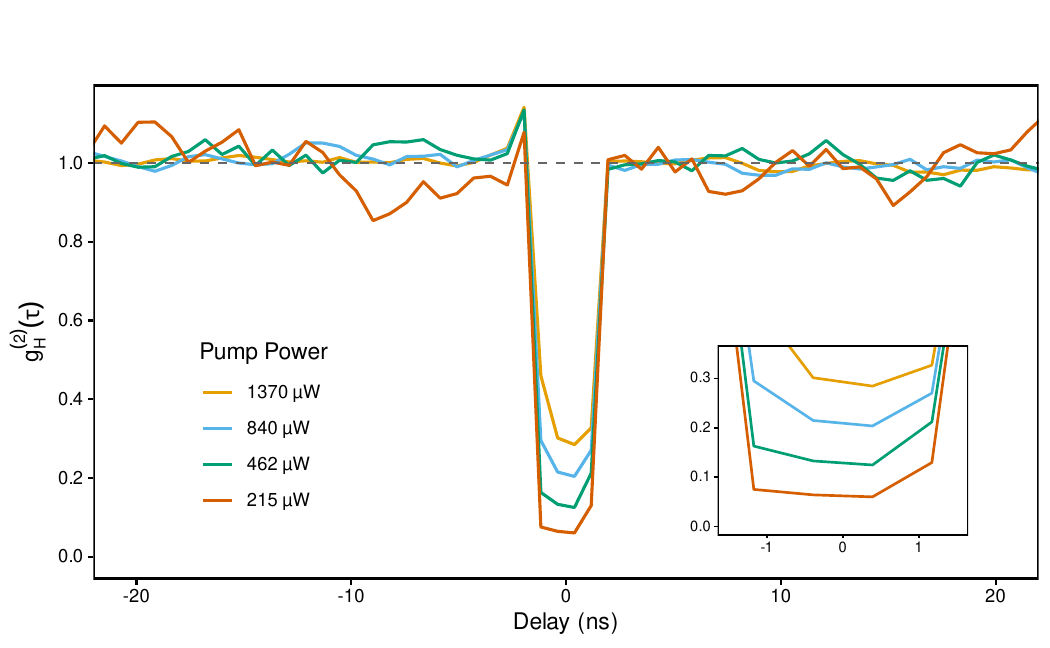}
    \caption{Heralded second-order correlation function measured at different pump powers. At the lowest pump power ($\qty{215}{\micro\watt}$), we measure $g^{(2)}_\mathrm{H}(0)\approx0.06$. The increase in $g^{(2)}_\mathrm{H}(0)$ with pump power is consistent with enhanced multi-pair generation at higher SPDC brightness.}\label{fig:g2plot}
\end{figure}

In the experiment shown in Fig.~\ref{fig:g2_setup}, $g^{(2)}_\mathrm{H}(\tau)$ was determined from the measured singles and coincidence rates. The heralded second-order correlation function is given by~\cite{poitrasProposalLownoiseHeralded2018}:
\begin{align}
    g^{(2)}_\mathrm{H} = \frac{T_\mathrm{123}S_\mathrm{1}}{R_\mathrm{12}R_\mathrm{13}},
\end{align}
where $T_\mathrm{123}$ is the triple coincidence rate, $S_\mathrm{1}$ is the singles rate at the signal detector, and $R_\mathrm{12}$ and $R_\mathrm{13}$ are the twofold coincidence rates between the signal detector and each of the idler detectors. The results are shown as a function of pump power in Fig.~\ref{fig:g2plot}.

\subsection{Spectral characterization}\label{section:spectral}

\begin{figure}[t]
    \centering\includegraphics[width=0.95\textwidth]{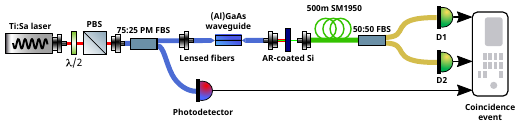}
    \caption{Experimental setup for time-of-flight spectral characterization of the generated photon pairs. Pulses from a mode-locked Ti:sapphire laser centered at \qty{980}{\nano\meter} with a repetition rate of \qty{80}{\mega\hertz} are coupled into the waveguide using polarization-maintaining single-mode lensed fibers. Phase-matching is achieved by tuning the center wavelength and bandwidth of the pump with a pulse shaper (not shown). The shaped pump has a bandwidth of approximately \qty{0.5}{\nano\meter} (FWHM). Residual pump light is suppressed by a free-space propagation region containing a half-wave plate, a polarizer and an AR-coated silicon window. The filtered beam is coupled into a \qty{500}{\meter} long SM1950 fiber followed by a 50:50 fiber beamsplitter (FBS). The photon pairs are detected with SNSPDs optimized for \qty{2}{\micro\meter} wavelengths. Arrival times are recorded by a HydraHarp 400 in T3 mode to reconstruct the joint spectrum.}\label{fig:exp}
\end{figure}

To quantify the spectral correlations, we evaluate the JSI, which measures frequency correlations between signal and idler photons. This measurement is challenging at \qty{2}{\micro\meter} primarily because of the scarcity of efficient single-photon detectors in this band. Scanning narrowband filters or monochromators can be used to measure a JSI at any wavelength, but they discard most of the photon flux and therefore require long acquisition times for faint sources. 

Instead, we use fiber-based time-of-flight spectroscopy~\cite{zielnickiJointSpectralCharacterization2018}, which maps wavelength onto arrival time after propagation through a long dispersive fiber. This technique detects the full spectrum simultaneously without blocking frequency components, with higher sensitivity and signal-to-noise ratio than scanning methods.

We measured the JSI by recording the time delay between a pump pulse and photon-pair detection after propagation through the dispersive fiber. In the experiment shown in Fig.~\ref{fig:exp}, pump pulses from the mode-locked Ti:sapphire laser were time-tagged together with detection events from the two SNSPD channels, and coincidence histograms were converted to a joint spectrum using a calibration of the fiber dispersion. 

To determine the mapping between arrival time and wavelength, we coupled a pulsed \qty{1550}{\nano\meter} laser into a highly nonlinear fiber to generate a supercontinuum, which was then injected into the spectrometer. Using reference bandpass filters centered at \qty{1900}{\nano\meter} and \qty{2000}{\nano\meter}, together with the known wavelength of the Ti:sapphire pump, we obtained three reference arrival times and performed a linear fit, which we then used to convert raw coincidence time delays into a joint spectrum.

The resulting JSIs for GaAs and AlGaAs, shown in Fig.~\ref{fig:JSIs}, indicate that the pairs are centered at the \qty{1960}{\nano\meter} degeneracy and span approximately \qty{200}{\nano\meter}. In both materials, we observe tight energy-conserving frequency anti-correlations.

\section{Numerical modeling}\label{section:numerical}

To understand device performance and inform future source design, we develop a numerical model of photon-pair generation in the waveguide. The model provides both the biphoton wavefunction (BWF) and the probability of generating a photon pair that survives propagation through the device.

Propagation loss is incorporated by assigning each guided mode $j$ a complex wavevector $\tilde{k}_j = k_j + i\kappa_j$, where $k_j$ is the usual wavevector for mode $j$, and $\kappa_j$ is an absorption parameter modeling propagation losses \cite{banicTwoStrategiesModeling2022}. In this work, the loss parameters $\kappa_j$ are derived from measured propagation losses in the device \cite{ulsigEfficientWidelyTunable2024}, given in Table~\ref{tab:Params}. We retain the frequency dependence of $k_j(\omega)$ to accurately model light generated over a large bandwidth, where guided-mode dispersion cannot be neglected.

This loss model applies strictly to cases where both down-converted photons reach the output and neither is scattered out of the waveguide~\cite{banicTwoStrategiesModeling2022}. Consequently, the complex wavevector model does not account for the statistics of broken pairs, that is, generation events where at least one photon is scattered and escapes coincidence detection. Despite this limitation, the model reproduces the measured joint spectra and provides useful estimates of pair-generation probabilities for lossy waveguides.

\subsection{Theoretical model of SPDC in lossy waveguides}

For sufficiently low pump powers, such that at most a single photon pair is generated per pump pulse, we approximate the state consisting of vacuum and a surviving photon pair as \cite{quesadaPhotonPairsNonlinear2022}:
\begin{equation}
    \ket{\Psi_{\mathrm{eff}}} \approx \ket{\mathrm{vac}} + \sqrt{\mathcal{P}} \ket{\mathrm{II}},
\end{equation}
where $\mathcal{P}$ is the probability of generating a photon pair that survives transit through the waveguide for a given pump pulse, obtained numerically by normalizing the BWF. To compare the model to experimental results, we define a down-conversion efficiency $P_{\mathrm{SPDC}}$ similar to Eq. \eqref{eq:P_SPDC}:
\begin{align}
    P_{\mathrm{SPDC}}^{\mathrm{pulse}} \equiv \mathcal{P} / N_{\mathrm{pump}},
    \label{eq:ProbPulsed}
\end{align}
which is the probability that one of the $N_{\mathrm{pump}}$ photons in the pump pulse is down-converted to a photon pair.

The normalized two-photon state is written in terms of a BWF:
\begin{equation}
    \ket{\mathrm{II}} = \frac{1}{\sqrt{2}} \iint \varphi_{\mathrm{II}}(\omega_s,\omega_i) a^\dagger(\omega_s)a^\dagger(\omega_i)\ket{\mathrm{vac}}
    \, \mathrm{d}\omega_s \,\mathrm{d}\omega_i   ,
\end{equation}
where $a(\omega)$ and $a^\dagger(\omega)$ are bosonic operators with the commutation relations $[a(\omega),a^\dagger(\omega')]=\delta(\omega-\omega')$.

From an asymptotic field approach \cite{banicTwoStrategiesModeling2022,liscidiniAsymptoticFieldsHamiltonian2012}, we obtain the normalized BWF:
\begin{equation}
    \varphi_{\mathrm{II}}(\omega_s,\omega_i) = \sqrt{\frac{E_p\omega_s\omega_i(\omega_s+\omega_i)}{8\mathcal{P}\pi\omega_p v_{s/i}(\omega_s)v_{s/i}(\omega_i)v_p(\omega_s+\omega_i)}} K(\omega_s+\omega_i, \omega_s, \omega_i) \phi(\omega_s+\omega_i),
\end{equation}
where $\phi(\omega)$ is an $L^2$ normalized function that models the spectral profile of the pump pulse, $E_p$ is the energy in each incident pump pulse with center frequency $\omega_p$, and $v_p(\omega)$ and $v_{s/i}(\omega)$ are the group velocities of the guided pump and signal/idler modes, respectively,
evaluated at frequency $\omega$. The nonlinear interaction strength of the fields in a waveguide of length $L$ can be characterized by a parameter $K$: 
\begin{equation}
\begin{split}
K(\omega_p,\omega_s,\omega_i)  = \frac{ \bar{\chi}^{(2)} L}{\epsilon_0^{1/2}\bar{n}^3\sqrt{A_{\mathrm{eff}}}} e^{\frac{i}{2}(\Delta k + \frac{i}{2}\Delta\kappa) L} e^{-\frac{1}{2}(\kappa_s(\omega_s) + \kappa_i(\omega_i)) L}  \times \mathrm{sinc}\left[ \left(\Delta k + \frac{i}{2}\Delta \kappa\right)\frac{L}{2} \right],
\end{split}
    \label{eq:Tensor3}
\end{equation}
where $\bar{\chi}^{(2)}$ is the nominal value of the second-order nonlinear susceptibility (taken as 238~pm/V for GaAs and 210~pm/V for AlGaAs \cite{ulsigEfficientWidelyTunable2024}), with $\mathrm{sinc}(x)\equiv \sin(x)/x$.
We define the effective area~\cite{fontainePhotonpairGenerationDownconversion2025}:
\begin{equation}
    A_{\mathrm{eff}} \equiv \frac{{N}_s{N}_i{N}_p}{\left|\iint_{\mathrm{wg}} \frac{\chi^{(2)}_{lmn}}{\bar{\chi}^{(2)}}\left[{\mathsf{e}}_{{s}}^{l}(x,y) {\mathsf{e}}_{{i}}^{m}(x,y)\right]^\ast {\mathsf{e}}_{{p}}^{n}(x,y)
    \,\mbox{d} x \,\mbox{d} y
    \right|^2},
    \label{eq:EffArea}
\end{equation}
with:
\begin{equation}
    N_j \equiv \frac{c}{\bar{n}^2}\iint \frac{n(\omega_j;x,y)}{v_{\mathrm{g}}(\omega_j;x,y)} \vec{\mathsf{e}}_j^*(x,y)  \cdot \vec{\mathsf{e}}_j(x,y)
    \,\mbox{d} x \,\mbox{d} y .
    \label{eq:N_j}
\end{equation}
Here, $n$ and $v_{\mathrm{g}}$ denote the material refractive index and material group velocity, respectively, evaluated from the Weber model~\cite{weberPropagationLightPeriodic1990}. The material group velocity in Eq.~\eqref{eq:N_j} is distinct from the guided-mode group velocities $v_p$ and $v_{s/i}$ used in Eqs.~\eqref{eq:Tensor3} and \eqref{eq:Rate_PairsOmega2_TLDR}. The explicit dependence on $(x,y)$ accounts for the spatial variation of the material dispersion across the simulation window. 
The effective area describes the modal overlap between the TE$_{00}$ and TM$_{00}$ modes, and Eq. \eqref{eq:EffArea} includes the full tensor nature of the $\chi^{(2)}$ interaction, as well as the full vector electric fields, where the Cartesian indices $lmn$ are implicitly summed. 
The normalization factors ${N}_j$ ensure that the electric-field profiles are properly normalized.
We have defined $\Delta k \equiv k_p(\omega_p) - k_s(\omega_s) - k_i(\omega_i)$ and $\Delta \kappa = \kappa_p(\omega_p) - \kappa_s(\omega_s) - \kappa_i(\omega_i)$, where $k_p(\omega)$ and $k_{s/i}(\omega)$ encode the dispersion relations for the TM$_{00}$ pump mode and TE$_{00}$ signal--idler modes, respectively, and similarly for the loss parameters $\kappa_p$, $\kappa_s$, and $\kappa_i$.

In the limit of continuous-wave (CW) excitation, we obtain an explicit expression for the generation rate for an input on-chip power of $P_{\mathrm{in}}$ at frequency $\omega_{\mathrm{p}}$:
\begin{equation}
\begin{split}
    {R_{\mathrm{cw}} = \frac{ P_{\mathrm{in}}}{8\pi v_p(\omega_{{p}})}
    \int \frac{  \omega_{{s}} (\omega_{{p}}
    -\omega_{{s}}) }
    {v_{s/i}(\omega_{\mathrm{s}}) v_{s/i}(\omega_{{p}}-\omega_{{s}})}  |K(\omega_{{p}},\omega_{{s}},\omega_{{p}}-\omega_{{s}})|^2
    \,\mathrm{d} \omega_{{s}}  .}
    \end{split}
    \label{eq:Rate_PairsOmega2_TLDR}
\end{equation}
Similar to the pulsed case, we define the SPDC efficiency as:
\begin{equation}
    P_{\mathrm{SPDC}}^{\mathrm{cw}} \equiv R_{\mathrm{cw}} \hbar \omega_p / P_{\mathrm{in}}.
    \label{eq:ProbCW}
\end{equation}

To model a measurement in which the pump is detuned from the phase-matching wavelength, we evaluate Eq.~\eqref{eq:Rate_PairsOmega2_TLDR} at the detuned pump frequency $\omega_{p}$, with no other model parameters changed; the reduction in efficiency then follows directly from the phase mismatch $\Delta k$ at the new pump wavelength.

\subsection{Simulations and comparisons}
To compare model predictions with experimental data, we first extract the optical properties of the guided modes using a finite-difference mode solver (EMode~\cite{EMode}) with the same material dispersion model. We determine the nonlinear interaction strength by calculating the modal overlap of the interacting fields: two TE$_{00}$ modes at \qty{1960}{\nano\meter} for the signal and idler, and a TM$_{00}$ mode at \qty{980}{\nano\meter} for the pump. 

This overlap yields effective areas $A_{\mathrm{eff}}$ of $0.759\,\mu$m$^2$ for GaAs and $0.749\,\mu$m$^2$ for AlGaAs. The main components of the mode profiles used in this calculation are shown in Fig.~\ref{fig:ModeFields}(a,b). We also extract the wavelength-dependent effective indices and group velocities required to evaluate the biphoton wavefunction from parameter sweeps of the corresponding modes. The parameters used in our simulations are summarized in Table~\ref{tab:Params}.

\begin{figure}[tb]
\centering
{
\includegraphics[width=0.95\linewidth]{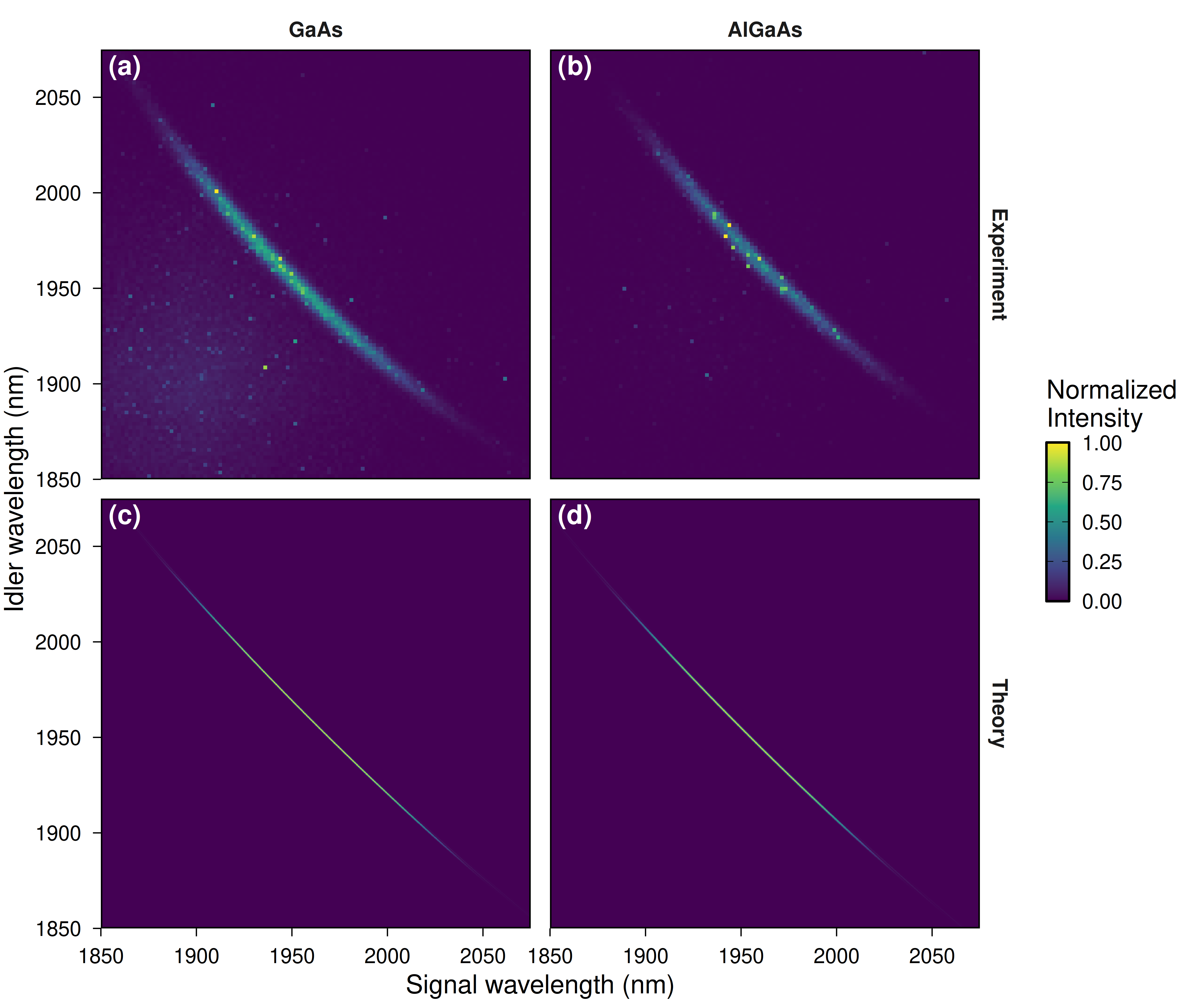}}
\centering
\caption{Joint-spectral intensities obtained from the time-of-flight experiment described in Section~\ref{section:spectral} for (a) GaAs and (b) AlGaAs. Simulated joint-spectral intensities, $\mathrm{JSI}(\lambda_s,\lambda_i)\propto |\varphi_{\mathrm{II}}(\lambda_s,\lambda_i)|^2$,
are shown
for (c) GaAs and (d) AlGaAs.}
\label{fig:JSIs}
\end{figure}

Fig.~\ref{fig:JSIs} compares the measured (a,b) and simulated (c,d) JSIs, showing that the numerical model reproduces the broadband structure of the generated photon pairs. The simulations use the nominal design geometry of each device, with heights of \qty{151}{\nano\meter} for GaAs and \qty{152}{\nano\meter} for AlGaAs~\cite{ulsigEfficientWidelyTunable2024}, with no tuned parameters. Experimentally, the pump wavelength reported in Table~\ref{tab:Params} corresponds to the center wavelength of the Ti:sapphire laser tuned to maximize the measured coincidence rate, which identifies the phase-matching condition of each device. 

In the simulations, we choose the pump wavelength $\lambda_p$ where the effective indices of the TE$_{00}$ and TM$_{00}$ modes intersect near \qty{1960}{\nano\meter} and \qty{980}{\nano\meter}, respectively. The simulated crossings reproduce the measured phase-matching wavelengths to within \qty{0.9}{\nano\meter}, corresponding to a waveguide-height deviation below \qty{0.2}{\nano\meter}. Since the devices were designed for modal phase matching with the same solver~\cite{ulsigEfficientWidelyTunable2024}, this agreement indicates that the fabricated waveguides closely matched the intended design. 

The anti-diagonal width of the measured JSIs exceeds that of the simulations because the measurement is limited by the spectral resolution of the time-of-flight spectrometer. The detector timing jitter of \qty{200}{\pico\second}, combined with the calibrated fiber dispersion of \qty{61}{\nano\meter\per\nano\second}, corresponds to a resolution of \qty{12}{\nano\meter} per photon, whereas the simulated width is set by the \qty{0.5}{\nano\meter} pump bandwidth.

While the model shows close agreement with the broad spectral features of the source, the predicted and measured absolute generation efficiencies in Table~\ref{tab:Params} differ. For AlGaAs, the continuous-wave measurement used a fixed-wavelength pump that was detuned by $\delta = 1.8 \pm 0.3$~\unit{\nano\meter} below the device's phase-matching wavelength, as the tunable pulsed pump was not available for that measurement (Section~\ref{subsec:efficiency}).
  
The model predicts a suppression to $0.37$ of the phase-matched efficiency at $\delta = 1.8$~nm. To compare with the measurement, we evaluate the model with the pump detuned by $\delta$ from its simulated phase-matching wavelength. Accounting for this detuning brings the predicted efficiency substantially closer to the measurement, reducing the discrepancy to a factor of $\approx2.6$, although the predicted and measured uncertainty ranges do not quite overlap. For GaAs, measured with a phase-matched pump, a comparable discrepancy of a factor $\approx2.8$ remains. 

These residual discrepancies can be attributed to uncertainty in determining the on-chip pump power from cutback measurements. The cutback estimate is obtained from separate paperclip test structures whose couplers are physically distinct from those of the device used for the pair-generation measurement. Since the chips were repeatedly handled between measurements, local damage to the couplers of the measured device, not captured by the cutback estimate, may make its true coupling loss higher than inferred. This would lead to an overestimated pump power and an artificially low experimental efficiency.
  
The uncertainties quoted for $P_{\mathrm{SPDC}}$ in Table~\ref{tab:Params} are obtained as follows. For the simulated values, the pump and signal--idler propagation losses are varied together to the limits of their measured uncertainties, which bounds the efficiency since it decreases monotonically with both losses; for AlGaAs, this range is combined in quadrature with the change in efficiency produced by the $\pm0.3$~nm uncertainty on the pump detuning. For the experimental values, the uncertainty is dominated by the pump coupling loss used to infer the on-chip pump power from the incident power.

\begin{table}[ht]
\centering
\renewcommand{\arraystretch}{1.3}
\begin{tabular}{@{} c | c c | c c @{}}
\hline\hline
& \multicolumn{2}{c|}{GaAs} & \multicolumn{2}{c}{AlGaAs} \\
\cline{2-5}
 & Experiment & Simulation & Experiment & Simulation \\
\hline
\makecell[c]{Pump wavelength \\ $\lambda_{\mathrm{p}}$ [\unit{\nano\meter}]} & 980.2 & 979.8 & 977.1 & 976.2 \\
\hline
\makecell[c]{Waveguide dimensions \\ (height$\times$width) [\unit{\nano\meter\squared}]} & (151$\times$1758) & (151$\times$1758) & (152$\times$1500) & (152$\times$1500) \\
\hline
Waveguide length [\unit{\micro\meter}] & 1678 & 1678 & 1678 & 1678 \\
\hline
Pump loss $\alpha_{\mathrm{p}}$ [dB/cm] & $18.1\pm5.7$ & $18.1\pm5.7$ & $40.0\pm6.1$ & $40.0\pm6.1$ \\
\hline
\makecell[c]{Signal/idler loss \\ $\alpha_{s/i}$ [dB/cm]} & $1.75\pm0.61$ & $1.75\pm0.61$ & $2.5\pm1.0$ & $2.5\pm1.0$ \\
\hline
\makecell[c]{Pump coupling loss \\ per facet [dB]} & $20.0\pm4.5$ & --- & $23.3\pm3.1$ & --- \\
\hline
$P_{\mathrm{SPDC}}$ ($\times 10^{-6}$) & $1.77^{+2.11}_{-0.98}$ & $5.04^{+0.34}_{-0.32}$ & $0.51^{+0.61}_{-0.28}\,^\dagger$ & $1.32^{+0.13}_{-0.16}\,^\dagger$ \\
\hline\hline
\end{tabular}
\caption{Experimental and simulated parameters used to determine JSIs and generation probabilities. The simulations use the nominal design heights~\cite{ulsigEfficientWidelyTunable2024} with no tuned parameters, reproducing the measured phase-matching wavelengths to within \qty{0.9}{\nano\meter}. The generation probabilities are calculated using the theoretical models corresponding to their respective experimental pump conditions: a pulsed pump for GaAs using Eq.~\eqref{eq:ProbPulsed} at the phase-matched pump wavelength, and a continuous-wave pump for AlGaAs using Eq.~\eqref{eq:ProbCW}. Simulated $P_{\mathrm{SPDC}}$ uncertainties are obtained by propagating the measured uncertainties on the pump and signal--idler propagation losses. Entries marked $^\dagger$ correspond to a pump detuned by $\delta=1.8\pm0.3$~nm below the phase-matching wavelength: the continuous-wave AlGaAs measurement was performed with the pump fixed at \qty{975.3}{\nano\meter}, and the simulated value is evaluated at the same detuning from its own phase-matching point, with the uncertainty on $\delta$ included in quadrature. At the phase-matching wavelength, the simulated AlGaAs efficiency is approximately $3.6\times10^{-6}$.}
\label{tab:Params}
\end{table}

\subsection{Optimization of waveguide length}

Despite the remaining discrepancy in absolute efficiency for GaAs, the model captures the underlying scaling and can guide future device designs. As an example, we optimize the waveguide length for maximum source brightness. The achievable surviving-pair probability is set by a trade-off between nonlinear interaction length and propagation loss. As shown in Table~\ref{tab:Params}, the pump mode at \qty{980}{\nano\meter} has much higher loss than the signal and idler modes at \qty{1960}{\nano\meter}.

\begin{figure}[tb]
\centering
{
\includegraphics[width=0.95\linewidth]{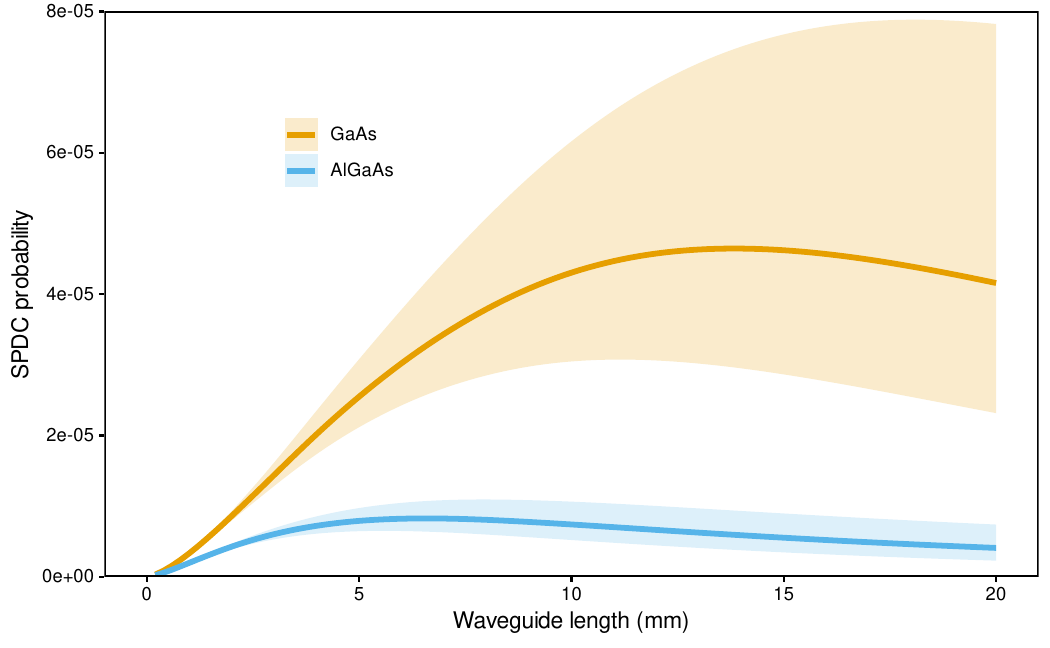}}
\centering
\caption{Predicted surviving-pair probability (SPDC probability per pump photon) as a function of waveguide length for GaAs and AlGaAs waveguides. The shaded bands are obtained by propagating the measured uncertainties in the pump and signal--idler propagation losses. The competition between nonlinear gain and pump attenuation yields an optimal waveguide length of approximately $13.8$ mm for GaAs and $6.6$ mm for AlGaAs.}
\label{fig:rates}
\end{figure}

In Fig.~\ref{fig:rates} we plot the predicted surviving-pair probability as a function of waveguide length, with simulations assuming a CW pump for both GaAs and AlGaAs devices. For short waveguides, this probability increases with length as nonlinear interaction dominates. However, as length increases, high pump loss begins to suppress efficiency. The simulation indicates an optimal length of approximately $13.8$ mm for GaAs and $6.6$ mm for AlGaAs for the current loss parameters, corresponding to an expected increase in surviving-pair probability by a factor of $6.8$ and $2.3$, respectively, compared to the $1.678$ mm devices characterized here. While longer waveguides are typically preferred for SPDC, high pump propagation loss in these ridge waveguides imposes a practical limit. Reducing propagation loss through improved fabrication would allow longer optimal interaction lengths, whereas shorter devices may be preferable for relaxing phase-matching constraints and minimizing wafer footprint.

\section{Discussion and conclusions}

In this work, we characterize infrared SPDC in GaAs- and AlGaAs-on-insulator waveguides. The sources exhibit low multi-pair probability and high coincidence-to-accidentals ratio, despite the high propagation losses at the pump wavelength.

By incorporating measured propagation losses and wavelength-dependent dispersion, our SPDC model reproduces the observed broadband emission spectra. Once the detuning of the continuous-wave pump from the phase-matching wavelength is taken into account, the predicted AlGaAs pair-generation efficiency is brought close to the measurement, reducing the discrepancy to a factor of $\approx2.6$; a comparable factor of $\approx2.8$ remains for GaAs. We believe these residual discrepancies are primarily associated with uncertainty in the estimated on-chip pump power derived from cutback measurements, rather than limitations of the underlying physical model; the framework therefore remains useful for future design optimization.

Among other integrated $\chi^{(2)}$ platforms, \ce{InGaP}-on-insulator has been used for SPDC at telecom wavelengths~\cite{akinInGaPH2Integrated2024,thielWaferscaleFabricationInGaPoninsulator2024,ahlerSecondorderNonlinearFrequency2025}, though recent broadband characterization shows a sharp increase in propagation loss between \qty{1950}{\nano\meter} and \qty{2325}{\nano\meter} \cite{ahlerLowlossInGaPoninsulatorWaveguides2026a}. By comparison, the (Al)GaAs-on-insulator devices demonstrated here have signal--idler propagation losses of
1.75 dB/cm to 2.5 dB/cm
near \qty{1960}{\nano\meter} (Table~\ref{tab:Params}).

The measured joint spectra reveal broadband, strongly anti-correlated photon-pair emission, in agreement with the phase-matching bandwidth predicted by the numerical model. Although the broad bandwidth increases the total pair rate, the resulting signal--idler spectral correlations reduce heralded single-photon purity and limit Hong--Ou--Mandel visibility unless narrowband filtering is applied or the joint spectral amplitude is engineered to be factorable~\cite{griceEliminatingFrequencySpacetime2001,mosleyHeraldedGenerationUltrafast2008,ansariHeraldedGenerationHighpurity2018}. On the other hand, applications such as photon precertification \cite{cabelloLoopholeFreeBellTest2012,meyer-scottCertifyingPresencePhotonic2016} place much weaker constraints on the spectral properties of the emitted photons and instead prioritize down-conversion efficiency, so the broadband emission of these waveguides may not represent a limitation for such applications.

The broadband and strongly time-frequency-entangled states generated in our devices could be advantageous for protocols that exploit this spectral structure. In particular, they are relevant to quantum spectroscopy and metrology~\cite{kalashnikovInfraredSpectroscopyVisible2016,pirandolaAdvancesPhotonicQuantum2018}, and to sensing across many spectral modes simultaneously, including time- and frequency-resolved Raman and fluorescence spectroscopies~\cite{zhangEntangledPhotonsEnabled2022}.

Heterogeneous bonding of (Al)GaAs-on-insulator to silicon-on-insulator, recently demonstrated at telecom wavelengths~\cite{lazzariBiphotonStateGeneration2026}, could bring these sources alongside maturing infrared silicon photonic components~\cite{linMidinfraredIntegratedPhotonics2018}. A complementary direction is to incorporate the pump laser on chip, building on recent demonstrations of heterogeneously integrated lasers at \qty{980}{\nano\meter}~\cite{naderHeterogeneousTantalaPhotonic2025}. Together, these advances provide a path toward fully integrated photon-pair sources near \qty{2}{\micro\meter}.

\section{Funding}
A.Z.L. and D.R.H. acknowledge the support of the Natural Sciences and Engineering Research Council of Canada, the Canada Foundation for Innovation, Canada Research Chairs, and the New Brunswick Innovation Foundation. S.E.F. and J.E.S. acknowledge the Natural Sciences and Engineering Research Council of Canada, and the European Union's Horizon Europe Research and Innovation Programme (101070700, project MIRAQLS) for financial support. S.E.F. acknowledges support from a Walter C. Sumner Memorial Fellowship.
N.V. and E.Z.U. acknowledge support from the Independent Research Fund Denmark (DFF).
NIST work was funded solely by the United States Government.

\section{Acknowledgments}

We thank Ryan DeCrescent for assistance with the Ti:sapphire laser setup and Michael Sloan for helpful discussions.

\section{Disclosures}
Authors E.Z.U., E.J.S., and N.V. are members of QVersion. E.J.S. is also a member of EMode Photonix. The remaining authors declare that the research was conducted in the absence of any commercial or financial relationships that could be construed as a potential conflict of interest.

\begin{quote}
\small
\noindent\textbf{Author contributions}\par
\noindent A.Z.L., E.Z.U., E.J.S., D.V.R., M.J.S., L.K.S., R.P.M., N.V., and D.R.H. conceived the project. E.Z.U., E.J.S., and R.P.M. designed the structures and fabricated the devices. S.E.F. and J.E.S. developed the theoretical model and performed the numerical simulations, with contributions from E.Z.U., E.J.S., and N.V. A.Z.L., E.Z.U., E.J.S., and D.V.R. conducted the experiments. A.Z.L. and E.Z.U. performed the data analysis. A.Z.L. and S.E.F. wrote the manuscript with contributions from all authors. E.J.S., D.V.R., M.J.S., L.K.S., J.E.S., N.V., and D.R.H. supervised the project.
\end{quote}

\section*{Disclaimer}
This document has not been peer reviewed but has been cleared by NIST for release.
\section*{Data availability statement}
The data that support the findings of this study are available from the corresponding author upon reasonable request.

\bibliography{biblio}

\end{document}